\documentclass[english]{article}
\usepackage[T1]{fontenc}
\usepackage[latin1]{inputenc}
\usepackage{amsmath}
\usepackage{amssymb}

\makeatletter
\usepackage{amsfonts}
\newtheorem{theorem}{Theorem}

\newtheorem{lemma}[theorem]{Lemma}

\newtheorem{remark}[theorem]{Remark}

\usepackage{babel}
\makeatother
\begin{document}

\title{Microscopic Models for Chemical Thermodynamics}

\author{V. A. Malyshev %
\footnote{INRIA, Rocquencourt, BP 105, 78153, Le Chesnay Cedex, France. Email:
Vadim.Malyshev@inria.fr. Phone: +33 1 39635269. Fax: +33 1 39635372%
}}

\maketitle
\begin{abstract}
We introduce an infinite particle system dynamics, which includes
stochastic chemical kinetics models, the classical Kac model and free
space movement. We study energy redistribution between two energy
types (kinetic and chemical) in different time scales, similar to
energy redistribution in the living cell. One example is considered
in great detail, where the model provides main formulas of chemical
thermodynamics.

Keywords: chemical thermodynamics, infinite particle systems, Kac
model, chemical kinetics. 
\end{abstract}
\markboth{Models of Chemical Thermodynamics}{Models of Chemical Thermodynamics}

\newpage

\section{Introduction}

As it is well-known, thermodynamical functions and some formulas of
the classical thermodynamics can be deduced from Gibbs canonical or
grand canonical ensemble. However, all thermodynamics, even the heat
exchange, has deeply dynamical nature and demands extra dynamics (this
is even more true for the chemical thermodynamics). This extra dynamics
can be modelled in various ways, satisfying however strong restrictins.
For example, in some scaling limit for small time scales it should
give thermodynamic formulas. In this paper we give examples of such
\char`\"{}thermo\char`\"{} dynamics, which, from one side, generalize
the classical Kac \cite{Kac} model (for convergence to Boltzman equation)
and stochastic chemical kinetics processes (see \cite{Leo,McQua}).
From another side, it adds energy parameter to Streater's statistical
dynamics \cite{Str} and Othmer's complex reaction networks theory
\cite{Oth,GadLeeOth}.

We try to concert this dynamics with the chemical thermodynamics and
with energy redistribution between \char`\"{}heat\char`\"{} energy
and chemical energy. We model this with two time scales. The first
time scale corresponds to the \char`\"{}fast\char`\"{} dynamics,
which governs the equilibrium behaviour, conserves the number of molecules
of each type and brings the system quickly to equilibrium. In fact
we assume that infinitely quickly. Second time scale corresponds to
the \char`\"{}slow\char`\"{} dynamics, that does not conserve
the number of particles, governs non-equilibrium process travelling
along the manifold of equilibrium distributions. We model this dynamics
with mean field dynamics of the stochastic chemical kinetics (taking
into account energy redistribution).

We describe here the simplest possible case, which demonstrates conceptual
picture without entering complicated mathematical techniques. Although
the calculations in this paper do not meet any technical difficulties,
it was not easy for me to reconstruct mathematically the conceptual
picture corresponding to what is written in chemical textbooks. Obviously
this model allows generalizations in many directions, including very
technical. We discuss some of them in the last section.

The generalization of the Kac model goes roughly as follows. First,
Kac type models (where a molecule possesses only kinetic energy, or
velocity) are married with stochastic chemical kinetics, where normally
the molecules differ only by type. As a result the molecule becomes
characterized by a pair $(j,T)$, that is by the type $j$ and kinetic
energy $T$. Then one more energy parameter is added to this pair
- the chemical energy $K$. However, still there are no volume, pressure
and other thermodynamic functions. Instead of introducing them apriori,
\char`\"{}by hands\char`\"{}, we throw particles independently
into $\Lambda$ and define the simplest possible but natural dynamics:
velocities are defined by kinetic energies, and the particles move
with constant velocities in-between the jumps.

The plan of the paper is the following. In the next section we define
the model and show the existence of the limiting infinite particle
dynamics, consisting of the free movement in space and some non-linear
Markov process, corresponding to reactions. This dynamics, after some
scaling limit, leaves invariant some manifold $\mathfrak{M}_{0}$
in the space of probability measures. This manifold is defined by
finite number of parameters and consists of Gibbs equilibrium distributions
for the mixture of ideal gases. We give selfcontained exposition of
formulas for thermodynamic functions in section 3. In section 4 we
introduce \char`\"{}thermo\char`\"{} dynamics (macroscopic evolution
of thermodynamic parameters) and show that it is a deterministic dynamical
system on $\mathfrak{M}_{0}$. In section 5 we consider unimolecular
reactions and chemical thermodynamics laws for this case. Last section
is devoted to discussion of further problems.

I thank Christian Maes for kind attention, fruitful remarks and stimulating
discussions.

\section{Microdynamics}

\paragraph{Molecules}

We consider molecules as classical point particles with translational
and internal (for example, rotational and/or vibrational) degrees
of freedom. More exactly, first of all, any molecule is characterized
by its type $j$. Translational degrees of freedom are given by its
velocity $\vec{v}\in R^{3}$, coordinate $\vec{x}\in R^{3}$ and the
kinetic energy $T_{j}=\frac{m_{j}\vec{v}^{2}}{2}$ (with mass $m_{j}$).
Internal degrees of freedom are assumed to be of two kinds - fast
and slow. They are also given by some energy functionals $I_{j}(y_{j}),K_{j}(z_{j}),y_{j}\in\mathbf{I}_{j},z_{j}\in\mathbf{K}_{j}$
in the space $\mathbf{I}_{j}\times\mathbf{K}_{j}$ of internal degrees
of freedom. It is often assumed, see \cite{LanLif}, that the total
energy of the molecule is\[
E_{j}=T_{j}(v_{j})+I_{j}(y_{j})+K_{j}(z_{j})\]
 Thus, we assume that the degrees of freedom of the molecule can be
\char`\"{}fast\char`\"{} or \char`\"{}slow\char`\"{}. The
corresponding parts of the energy will also be called either \char`\"{}fast\char`\"{}
or \char`\"{}slow\char`\"{}: $T$ is always fast, and say $I$
is fast, and $K$ is slow. One of our goals is to introduce some models
of energy redistribution between fast and slow parts.

For notation purpose only, we mostly take $I_{j}=0$, unless otherwise
stated, and we assume further on that $K_{j}=K_{j}(z_{j})$ is constant,
depending only on $j$. $K_{j}$ can be thought roughly as the energy
of some chemical bonds in the $j$ type molecule.

\paragraph{Reactions}

There are corresponding fast and slow reactions. Fast reactions do
not touch slow parts, but slow reactions may change both slow and
fast energies, thus providing energy redistribution between heat and
chemical energy. We consider here the following reactions:

\begin{enumerate}
\item slow unary (unimolecular) reactions $A\rightarrow B$; 
\item slow binary reactions of any type $A+B\rightarrow C+D$; 
\item fast binary reactions of the type $A+B\rightarrow A+B$, which draw
the system towards equilibrium; 
\item fast process of heat exchange with the environment, with reactions
of the type $A+B\rightarrow A+B$, but where one of the molecules
is an outside molecule. 
\end{enumerate}
In any considered reaction the total energy conservation is assumed,
that is the sum of total energies in the left side is equal to the
sum of total energies in the right side of the reaction equation.

Now we proceed to rigorous definitions.

\subsection{Finite volume dynamics}

System in a finite volume $\Lambda$ is defined as follows. Finite
number $N$ of particles (molecules) $i=1,...,N$, each equipped with
degrees of freedom $j,T,K$, are thrown uniformly and independently
in the cube $\Lambda$. The dynamics consists of two processes: basic
Markov jump process $M_{N}(t)$, describing the evolution of internal
degrees of freedom, and free space movement which will be defined
for each trajectory of $M_{N}(t)$. The process $M_{N}(t)$ has states
$\left\{ (j_{i},T_{i}),i=1,...,N\right\} $ and will be defined on
the state space $\mathbb{I}=(\left\{ 1,...,J\right\} \times R_{+})^{N}$.
The particles in the process $M_{N}(t)$ are ordered, because in the
space dynamics they will have coordinates.

\paragraph{Unary (unimolecular) reactions}

Assume that $J(J-1)$ nonnegative functions (rates) \[
u_{jj_{1}}=u_{jj_{1}}(T),j\neq j_{1}\]
 of $T\in R_{+}$ be given. Dependence on $T$ is very important to
be compatible with energy redistribution: for example, the reaction
may not be possible for low reactant kinetic energies. The rates do
not depend on $i$ and are assumed bounded functions of $T$. Unimolecular
reactions $A\rightarrow B$ have been used to describe many biological
processes, for example, protein and RNA conformational transitions,
transcription and translation processes etc.

Consider finite continuous time homogeneous Markov chain with state
space $\mathbb{I}$, which is defined as follows: any particle $i$
having at time $t$ type $j=j(t)$ and kinetic energy $T=T(t)$ waits,
independently of the other particles, some random time $\tau$. This
random time is assumed to have exponential distribution with the rate
$u_{j}(T)=\sum_{j_{1}}u_{jj_{1}}(T)$. Then at time $t+\tau$ the
particle chooses with probability $p_{jj_{1}}=\frac{u_{jj_{1}}}{\sum_{j_{1}}u_{jj_{1}}}$
the type $j_{1}$ to perform the jump $j\rightarrow j_{1}$. However
the jump does occur iff $T+K_{j}-K_{j_{1}}\geq0$. If $T+K_{j}-K_{j_{1}}<0$
then nothing occurs and the process starts anew.

One could equivalently say it like this. Define the allowed set\[
A(j,T)=\left\{ j_{1}:T+K_{j}-K_{j_{1}}\geq0\right\} \]
 of types for the particle of type $j$ and having kinetic energy
$T$. Then such particle waits exponential time with the rate\[
u_{j}(T_{i})=\sum_{j_{1}\in A(j,T)}u_{jj_{1}}(T_{i})\]
 and then changes its type to $j_{1}$ with probability\[
\frac{u_{jj_{1}}(T_{i})}{\sum_{j_{1}\in A(j,T)}u_{jj_{1}}(T_{i})}\]

The transformation $j\rightarrow j_{1}$ of the particle $i$ at time
$t+\tau$ is accompanied \ by the energy redistribution $(T,K_{j})\rightarrow(T_{1}=T+K_{j}-K_{j_{1}},K_{j_{1}})$.
In other words, we assume energy conservation $T_{1}+K_{j_{1}}=T+K_{j}$.

We also assume that the velocity $v$ immediately after the jump $j\rightarrow j_{1}$
becomes uniformly distributed on the sphere of radius $T_{1}$. Note
that one could assume much less.

We could define the state space of the process $M_{N}(t)$ as the
sequence of arrays $X_{i}=\left\{ j_{i},\vec{v}_{i}\right\} ,i=1,...,N$.
Then $n_{j}$ is the number of $i$ such that $j_{i}=j$, $N=n_{1}+...+n_{J}$.
However, due to our agreement about velocities we can take instead
the state of a molecule as $X_{i}=\left\{ j_{i},T_{i}\right\} $.
Denote $H_{N}^{(u)}$ the generator of the $N$-particle process,
defined on some appropriate function space on $(\left\{ 1,...,J\right\} \times R_{+})^{N}$,
which we will not write down explicitely, because it is quite standard.

For example, for one particle we have continuous time Markov process
$M_{1}(t)$ with the state space $(j,T)$: the $T(t)$ component is
uniquely defined by initial conditions and by the sequence of type
transformations.

\paragraph{Slow binary reactions}

Here the Markov jump process is the following. On the time interval
$(t,t+dt)$ each (ordered) pair $(i,i^{\prime})$ of molecules, with
parameters $(j,T\,),(j^{\prime},T^{\prime})$ correspondingly, has
a \char`\"{}collision\char`\"{} with probability $\frac{1}{N}b_{jj^{\prime}}dt,b_{jj^{\prime}}=b_{jj^{\prime}}(T,T^{\prime})=b_{j^{\prime}j}$.
Then at the moment of collision the parameters of the particles $i,i^{\prime}$
at time $t+0$ become correspondingly $(j_{1},T_{1}),(j_{1}^{\prime},T_{1}^{\prime})$.
The distribution of the new parameters is defined by the conditional
densities \[
P^{(b)}(j_{1},T_{1},j_{1}^{\prime}|(j,T),(j^{\prime},T^{\prime}))\]
 that are defined for any $j,j^{\prime},T,T^{\prime},j_{1},T_{1},j_{1}^{\prime}$,
then $T_{1}^{\prime}=T+K+T^{\prime}+K^{\prime}-K_{1}-T_{1}-K_{1}^{\prime}$.
To justify this definition it is assumed that \[
P^{(b)}(j_{1},T_{1},j_{1}^{\prime}|(j,T),(j^{\prime},T^{\prime}))=0\]
 if $T+K+T^{\prime}+K^{\prime}-K_{1}-T_{1}-K_{1}^{\prime}<0$, and\[
P^{(b)}(j_{1},T_{1},j_{1}^{\prime}|(j,T),(j^{\prime},T^{\prime}))\geq0\]
 if $T+K+T^{\prime}+K^{\prime}-K_{1}-T_{1}-K_{1}^{\prime}\geq0$.
Moreover, for any $(j,T),(j^{\prime},T^{\prime})$\[
\sum_{j_{1},j_{1}^{\prime}}\int dT_{1}P^{(b)}(j_{1},T_{1},j_{1}^{\prime}|(j,T),(j^{\prime},T^{\prime}))=1\]

We assume the same agreement about new velocities, that is each of
them is distributed independently and uniformy on the corresponding
energy sphere.

Denote $H_{N}^{(b)}$ the corresponding generator, on the same function
space.

\paragraph{Fast binary reactions}

Their definition is similar to slow binary reaction process, but the
particles do not change types and slow energies, so only redistribution
of their kinetic energies occurs, that is $T,T^{\prime}\rightarrow T_{1},T_{1}^{\prime}$.
The agreement concerning velocities holds as above. We write the collision
probabilities as $\frac{1}{N}f_{jj^{\prime}}dt,f_{jj^{\prime}}=f_{j^{\prime}j}$,
we assume also that in the reaction $j,j^{\prime}\rightarrow j,j^{\prime}$
the energy conservation $T+T^{\prime}=T_{1}+T_{1}^{\prime}$ holds.
We assume that $f_{jj^{\prime}}$ do not depend on $T$ and $T^{\prime}$.
As for the conditional distribution $P^{(f)}(T_{1}|T,T^{\prime})$,
we will use the one defined below, similar to the example in \cite{FaMaPi}.

We define Kac type models\ as follows. Assume that there is a family
$M(a),0\leq a<\infty$, of distributions $\mu_{a}$ on $R_{+}$ with
the following property. Take two i.i.d. random variables $\xi_{1},\xi_{2}$
with the distribution $M(a)$. Then their sum $\xi=\xi_{1}+\xi_{2}$
has distribution $M(2a)$. We assume also that $a$ is the expectation
of the distribution $M(a)$. Denote $p(\xi_{1}|\xi)$ the conditional
density of $\xi_{1}$ given $\xi$, defined on the interval $[0,\xi]$.
We put\[
P^{(f)}(T_{1}|T,T^{\prime})=p(T_{1}|T+T^{\prime})\]
 and of course $T_{1}^{\prime}=T+T^{\prime}-T_{1}$. Denote the corresponding
generator $H_{N}^{(f)}$.

\paragraph{Heat transfer}

We model it similarly to the fast binary reactions, as random \char`\"{}collisions\char`\"{}
with outside molecules in an infinite bath, which is kept at constant
temperature $\beta$. The energy of each outside molecule is assumed
to have $\chi^{2}$ distribution with $3$ degrees of freedom and
with parameter $\beta$. More exactly, for each molecule $i$ there
is Poisson process with some rate $h$. Denote $t_{ik},k=1,2,...$,
its jump moments, when it undergoes collisions with outside molecules.
At this moments the kinetic energy $T$ of the molecule $i$ is transformed
as follows. The new kinetic energy $T_{1}$ after transformation is
chosen correspondingly to conditional density $p$ on the interval
$[0,T+\xi_{ik}]$, where $\xi_{ik}$ are i.i.d. random variables having
$\chi^{2}$ distribution with density $cx^{\frac{1}{2}}\exp(-\beta x)$.
Denote the corresponding conditional density by $P^{(\beta)}(T_{1}|T)$.
In fact, this process amounts to $N$ independent one-particle processes,
denote the corresponding generator $H_{N}^{(\beta)}$.

\paragraph{Full dynamics}

Note that both for unary and slow binary reactions the numbers $n_{j}$
of type $j$ molecules are not conserved but the total number of molecules
$N=\sum_{j}n_{j}$ is conserved. On the contrary, fast reactions conserve
$n_{j}$. The process $M_{N}(t)$ is defined by the sum of generators\[
H(s_{f},s_{\beta})=H_{N}^{(u)}+H_{N}^{(b)}+s_{f}H_{N}^{(f)}+s_{\beta}H_{N}^{(\beta)}\]
 on some appropriate function space on $(\left\{ 1,...,J\right\} \times R_{+})^{N}$,
where $s_{f},s_{\beta}$ are some large scaling factors, which eventually
will tend to infinity. This process \ belongs to a class of well
studied classical processes.

The state space of the full process is the sequence of arrays $X_{i}=\left\{ j_{i},\vec{x}_{i},\vec{v}_{i}\right\} ,i=1,...,N$,
then $n_{j}$ is the number of $i$ such that $j_{i}=j$ molecules,
$N=n_{1}+...+n_{J}$. However, due to our agreement about velocities.
we can take instead the state of a molecule as $X_{i}=\left\{ j_{i},\vec{x}_{i},T_{i}\right\} $.

For each trajectory $\omega$ of $M_{N}(t)$ we define the local space
dynamics as follows. \ It is quite simple: it does not change types,
energies, velocities, but only coordinates. If at jump moment $t$
the particle acquires velocity $\vec{v}(\omega)=\vec{v}(t+0,\omega)$
and has coordinate $\vec{x}(t,\omega)$, then at time $t+s$\begin{equation}
\vec{x}(t+s,\omega)=\vec{x}(t,\omega)+\vec{v}(\omega)s\label{trans}\end{equation}
 unless the next event (jump), concerning this particle, of the trajectory
$\omega$ occurs on the time interval $\left[t,t+s\right]$. We assume
periodic boundary conditions, or, that is the same, elastic reflection
from the boundary.

We denote the resulting process $\mathfrak{X}_{\Lambda,N}(t)$. It
depends of course also on the initial conditions and on $s_{f},s_{\beta}$.

\begin{remark}
One of the disadvantages of the dynamics defined above is that there is no
momentum conservation. This could be justified by including a fast process
(similar to the defined above) of elastic collisions with some other smaller
particles, which quickly establishes uniform distriution of velocities\footnote{The author thanks one of the referees for this remark.}. Moreover, instead
of the motion with constant velocity in-between collisions in such models it
could be even more natural to consider isotropic diffusion \(w(t)\) with zero
drift and put\begin{equation*}
\vec{x}(t+s,\omega )=\vec{x}(t,\omega )+w(s)
\end{equation*}instead of (\ref{trans}). Instead of the velocity \(\vec{v}\) for one-particle
motion one would have the covariance \(\sigma ^{2}\) of the diffusion. One
should only fix somehow the dependence \(T(\sigma ^{2})\). Most considerations
below admit this generalization. Thius generalization could allow to take
into account deeper results concerning reaction-diffusion equations, see 
\cite{BraLeb, MaFeLe, ArnThe, Kote}.
\end{remark}

\subsection{Infinite particle dynamics}

We show that the infinite particle limit exists. However, we do not
get in general a Markov process, but only so called nonlinear Markov
process.

\paragraph{Unimolecular reactions}

Denote $p_{t}(j,T),j=1,...,J$, the densities (with respect to the
Lebesgue measure $dT$) of the one-particle process $M_{1}$ at time
$t$\[
\sum_{j=1}^{J}\int p_{t}(j,T)dT=1\]
 If there are only slow unary reactions, then due to the energy conservation,
for $M_{1}$ there exist (under mild conditions on $u_{jj_{1}}$)
the limiting densities $\pi(j,T)=\lim_{t\rightarrow\infty}p_{t}(j,T)$,
depending on the initial conditions, because this chain is strongly
reducible.

Define infinite particle dynamics $\mathfrak{X}_{p_{0}}(t)$ as the
collection of independent one-particle trajectories, where $p_{0}$
is the initial distribution for each particle. For infinite particle
case one particle trajectories are defined exactly as in a finite
volume, only there are no boundary conditions, and the particle moves
in the whole space.

Define the limiting concentrations $c_{j}(t)=\lim_{\Lambda\rightarrow\infty}\Lambda^{-1}<n_{j}(t>=p_{t}(j)c$,
where $p_{t}(j)$ is the probability that at time $t$ a molecule
has type $j$.

\paragraph{Heat transfer}

Heat transfer is also a one-particle process and in the infinite particle
evolution is defined similarly to the previous one. It will also be
Markov.

\paragraph{Binary reactions}

The case with binary reactions is more involved because one cannot
define mean field dynamics directly for infinite particle system.
Nevertheless, we define the so called nonlinear Markov process. It
consists of deterministic evolution of the densities $p_{t}(j,T)$,
defined by some Boltzman type equation, and infinite number of independent
inhomogeneous Markov jump processes for internal degrees of freedom
of the individual particles. These two evolutions are concerted with
each other in the sense we explain below. In this sense mean field
dynamics in the infinite limit becomes local. In other words, if we
are observing some local region, we never see simultaneous jumps of
two particles, but only jumps of one particle - the particle with
which it \char`\"{}collides\char`\"{} is a.s. infinitely far from
this region.

The exact definition proceeds quite similarly to the finite particle
dynamics with one essential difference. The infinite particle dynamics
consists of the process $M_{\infty}(t)$ and free space movement.
The jump process $M_{\infty}(t)$ is defined as collection of independent
one-particle processes $M_{\infty,1}$. Each of this one-particle
processes is defined by the following Kolmogorov equations\[
\frac{\partial p_{t}(j_{1},T_{1})}{\partial t}=\]
\begin{equation}
=\sum_{j}\int(P(t;j_{1},T_{1}|j,T)p_{t}(j,T)-P(t;j,T|j_{1},T_{1})p_{t}(j_{1},T_{1}))dT\label{kolbol}\end{equation}
 defining Markov process with distributions $p_{t}(j,T)$. To complete
the definition one should define the transition kernel. When all four
types of reactions are present, the kernel is\begin{equation}
P(t;j_{1},T_{1}|j,T)=\sum_{m=1}^{4}P^{(m)}(t;j_{1},T_{1}|j,T)\label{kernel}\end{equation}
 that is the sum of four terms $m=1,2,3,4\,$, corresponding to the
four reaction types, introduced above,\[
P^{(1)}=u_{jj_{1}}(T)\delta(T+K-K_{1}-T_{1}),\]
\begin{eqnarray*}
P^{(2)} & = & \sum_{j^{\prime},j_{1}^{\prime}}\int dT^{\prime}dT_{1}^{\prime}2b_{jj^{\prime}}P^{(b)}(j_{1},T_{1},j_{1}^{\prime}|(j,T),(j^{\prime},T^{\prime}))\\
 &  & p_{t}(j^{\prime},T^{\prime})\delta(T+K+T^{\prime}+K^{\prime}-K_{1}-T_{1}-K_{1}^{\prime}-T_{1}^{\prime}),\end{eqnarray*}
 \[
P^{(3)}=\delta_{jj_{1}}\sum_{j^{\prime}}\int P_{jj^{\prime}}^{(f)}(T_{1}|T,T^{\prime})2f_{jj^{\prime}}p_{t}(j^{\prime},T^{\prime})dT^{\prime},\]
 \[
P^{(4)}=h\delta_{jj_{1}}P^{(\beta)}(T_{1}|T)\]
 We see that terms 2 and 3 depend on \ $p_{t}(j,T)$. One should
find then $p_{t}(j,T)$ from equation, obtained by substituting (\ref{kernel})
into (\ref{kolbol}1\}). We get the Boltzman type equation.. In our
case it is not difficult to prove existence and uniqueness of $p_{t}(j,T)$,
because it is obtained as the limit of mean field processes. Also
the chaos property similar to one in the Kac model follows. The chaos
property corresponds to the independence of each particle jumps. Note
however that the processes $M_{\infty,1}$ are time inhomogeneous.

Consider now infinite particle system in $R^{3}$, where each particle
has internal parameters $(j,v)$. Denote $\mathfrak{M}$ the system
of all probability measures for this system with the following properties:

\begin{itemize}
\item coordinates of these particles are distributed as the homogeneous
Poisson point field of particles on $R^{3}$ with some density $c$, 
\item each particle has a vector of parameters $j,T$ distributed (independently
of \ its coordinate and of the other particles) via some common distribution
$p(j,T)$, the same for all particles. 
\end{itemize}
Consider a sequence of processes $\mathfrak{X}_{\Lambda,N}(t)$ with
$N=N(\Lambda),\Lambda\rightarrow\infty$ so that $\frac{N(\Lambda)}{\Lambda}\rightarrow c>0$.
Then at time $0$ the distribution of $\mathfrak{X}_{\Lambda,N}(0)$
converges to some distribution belonging to the set $\mathfrak{M}$.

\begin{theorem}
If the initial distribution belongs to \(\mathfrak{M}\) then the
infinite-particle dynamics exists, moreover \(\mathfrak{M}\) is invariant with
respect to this infinite particle dynamics. Under the conditions stated
above, the thermodynamic limit \(\mathfrak{X}_{c}(t)\) of the processes \(\mathfrak{X}_{\Lambda ,N}(t)\) exists and belongs to \(\mathfrak{M}\) at each
time moment \(t\).
\end{theorem}

Proof. To prove existence of the thermodynamic limit we proceed in
two steps. On the first step we do not care about coordinates. If
there are only unary reactions, there are no problem - one should
not perform the thermodynamic limit. But for binary reactions one
should.

Note that we can reformulate binary collisions in a finite volume
as follows. As an example we take fast binary collisions where we
can say that any particle $i$ of type $j$ undergoes \char`\"{}collision
with SOME molecule of SOME type $j^{\prime}$\char`\"{} with the
probability \[
dt\sum_{j^{\prime}}(f_{jj^{\prime}}+f_{j^{\prime}j})\frac{n_{j^{\prime}}(t)}{N}\]
 After the infinite volume limit the probability of collision becomes\begin{equation}
dt\sum_{j^{\prime}}(f_{jj^{\prime}}+f_{j^{\prime}j})\frac{c_{j^{\prime}}(t)}{c}\label{local}\end{equation}
 The same can be done for the energy distribution, that is we get
(\ref{kolbol} ).

To prove the existence of infinite particle dynamics one should prove
first the existence of solutions of the Boltzman equation, and then
the existence of the Markov process for the individual particle. All
these technicalities are put in the Appendix.

\begin{remark}
Note that in (\ref{local}) one could interpret \(c_{j^{\prime }}(t)\) as some
local concentration of particles in the vicinity of the \(j\) particle. This
gives some links to local dynamics. This hints on some possible
generalizations of the process which could include spatial correlations.
\end{remark}

On the next step, we need to prove that the homogeneous Poisson distribution
in space is invariant.

We use the following general lemma. Note that the assumption that
the velocities are uniformly distributed on the energy sphere is not
essential. It is sufficient only that the velocity process did not
depend on the coordinates.

\begin{lemma}
Let all particles in the configuration \(\left\{ x_{i}(t)\right\} \) in \(R^{d}\)
move (independently) as\begin{equation*}
x_{i}(t)=x_{i}(0)+\int v_{i}(s,\omega )ds
\end{equation*}where the velocities \(v_{i}(t,\omega )\) are independent processes with
arbitrary time dependence. Then the Poisson distribution is invariant.
\end{lemma}

Proof. Take $N$ such particles in the finite cube $\Lambda$ with
the same independent movement, only take periodic boundary conditions.
The result will follow if we prove that the uniform distribution on
$\Lambda^{N}$ is conserved with such dynamics. In fact, one can perform
afterwards the thermodynamic limit with $\frac{N}{\Lambda}\rightarrow c$.
Then the uniform distribution at any time $t$ converges to the Poisson
distribution, the finite volume dynamics at time $t$ converges to
the dynamics in $R^{d}$ because for fixed $t$ and sufficiently large
$\Lambda$ the particle does not reach the boundary of $\Lambda$.
Thus Poisson distribution is invariant. See also general results of
this kind in \cite{Dobr}.

Because of the independence it is sufficient to consider $N=1$ that
is one particle, thrown at time $0$ with uniform distribution into
$\Lambda$. Assume that the particle moves in this cube with some
speed $v(t)$ - arbitrary random function of time. The only condition
is that $v(t)$ does not depend on the coordinate of the particle.
Then it is clear that the uniform distribution is invariant.

\section{Thermodynamic functions for mixture of ideal gases}

In our case the Gibbs state will be the system of independent particles
(ideal gas). Here we give a selfcontained presentation (fixing the
notation we use here) of main formulas for the classical ideal gases
and mixtures, with one important difference: the fast degrees of freedom
are gaussian and slow degrees of freedom are constants $K_{j}$, depending
only on $j$.

A well-known example of internal energy functional is the quadratic
Hamiltonian \[
I_{j}=\sum_{k=1}^{d_{j}-3}\frac{m_{j,k}w_{j,k}^{2}}{2}\]
 where $m_{j,k},k=1,...,d_{j}-3$, are some coefficients and the vector
$y_{j}=\left\{ w_{j,k},k=1,...,d_{j}-3\right\} \in R^{d_{j}-3}$.
Then $d_{j}-3$ is the number of internal degrees of freedom of the
molecule of type $j$, $d_{j}$ is the number of all degrees of freedom.
This is justified, for example, when internal oscillations are small,
see \cite{LanLif}. We will call this the Gaussian case. We will need
another extreme case, when $K_{j}$ are constants.

Thus each molecule of type $j$ has the energy\[
E_{j}=T_{j}+I_{j}+K_{j}\]

We consider a finite number $n_{j}$ of particles of types $j=1,...,J$
in a finite volume $\Lambda$. For the ideal gas of the $j$ type
particles the grand partition function of the Gibbs distribution is\[
\Theta(j,\beta)=\sum_{n_{j}=0}^{\infty}\frac{1}{n_{j}!}\left(\prod\limits _{i=1}^{n_{j}}\int_{\Lambda}\int_{R^{3}}\int_{\mathbf{I}_{j}}d\vec{x}_{j,i}d\vec{v}_{j,i}dy_{j.i}\right) \times
\]
\[ \times \exp\beta(\mu_{j}n_{j}-\sum_{i=1}^{n_{j}}(\frac{m_{j}v_{j,i}^{2}}{2}+I_{j}(y_{j,i}))-K_{j})=\]
 \[
=\sum_{n_{j}=0}^{\infty}\frac{1}{n_{j}!}\Lambda^{n_{j}}\beta^{-\frac{d_{j}}{2}n_{j}}B_{j}^{n_{j}}\exp\beta(\mu_{j}-K_{j})n_{j}=\exp(\Lambda\beta_{j}^{-\frac{d_{j}}{2}}B_{j}\exp\beta_{j}\hat{\mu}_{j})\]
 where\[
B_{j}=(\frac{2\pi}{m_{j}})^{\frac{3}{2}}\prod\limits _{k=1}^{d_{j}-3}(\frac{2\pi}{m_{j,k}})^{\frac{1}{2}},\hat{\mu}_{j}=\mu_{j}-K_{j}\]
 General mixture distribution of $J$ types is defined by the following
partition function\[
\Theta=\prod\limits _{j=1}^{J}\Theta(j,\beta)=\exp(\Lambda\sum_{j}\lambda_{j}\exp\beta\hat{\mu}_{j}),\lambda_{j}=\beta^{-\frac{d_{j}}{2}}B_{j}\]
 Define the grand thermodynamic potential\[
\Omega=\Omega_{\Lambda}=-\beta^{-1}\ln\Theta=-\beta^{-1}\Lambda\sum_{j}\lambda_{j}\exp\beta\hat{\mu}_{j}\]
 The limiting space distribution of type $j$ particles is the Poisson
distribution with rate (concentration) $c_{j}$, and\[
c_{j}=\frac{<n_{j}>_{\Lambda}}{\Lambda}=\beta^{-1}\frac{\partial\ln\Theta}{\partial\mu_{j}}=\lambda_{j}\exp\beta\hat{\mu}_{j}=\exp(\beta\mu_{j}-\beta\mu_{j,0}-\beta K_{j})\]
 Put $c=c_{1}+...+c_{J}$. Then\begin{equation}
\mu_{j}=\beta^{-1}\ln(\frac{<n_{j}>}{\Lambda}\lambda_{j}^{-1})=\mu_{j,0}+\beta^{-1}\ln c_{j}+K_{j},\label{chempot}\end{equation}
 where\begin{equation}
\mu_{j,0}=-\beta^{-1}\ln\lambda_{j}=-\beta^{-1}(-\frac{d_{j}}{2}\ln\beta+\ln B_{j})\label{standard}\end{equation}
 is the so called standard chemical potential, it corresponds to the
unit concentration $c_{j}=1$.

The internal energy in thermodynamics is defined as the mean of the
sum of energies of all particles. The conditional mean energy (given
type $j$) particle is (the law of equipartition of energy) \[
<E_{j}>=\frac{d_{j}}{2}\beta^{-1}+K_{j}\]
 and\[
U=\sum_{j}<n_{j}>(\frac{d_{j}}{2}\beta^{-1}+K_{j})\]
 The pressure is defined as\[
P=-\frac{\partial\Omega}{\partial\Lambda}=\Lambda^{-1}\beta^{-1}\sum_{j}<n_{j}>=\beta^{-1}\sum_{j}c_{j}=\sum_{j}p_{j}\]
 where $p_{j}=\beta^{-1}c_{j}$ are the partial pressures. In the
thermodynamic limit this is equivalent to the definition \[
\beta P=\lim_{V\rightarrow\infty}\frac{1}{\Lambda}\ln\Theta\]
 The well-known equation of state follows\begin{equation}
P\Lambda=\beta^{-1}\sum_{j}<n_{j}>\label{eqstate}\end{equation}
 For one type $j$ the entropy is defined as\[
S_{j}=-\frac{\partial\Omega_{j}}{\partial(\beta^{-1})}=\Lambda\lambda_{j}\exp(\beta\hat{\mu}_{j})(\frac{d_{j}}{2}+1-\beta\hat{\mu}_{j})=<n_{j}>(\frac{d_{j}}{2}+1+\beta K_{j}-\beta\mu_{j})\]
 (Sackur-Tetrode formula). For the mixture it is the sum of these
$S=\sum_{j}S_{j}$.

Together with the internal energy $U$ three other important thermodynamic
potentials are: the enthalpy\begin{equation}
H=U+P\Lambda=\beta^{-1}\sum_{j}<n_{j}>(\frac{d_{j}}{2}+1+\beta K_{j}),\label{enth}\end{equation}
 Gibbs free energy \[
G=H-\beta^{-1}S=\sum_{j}\mu_{j}<n_{j}>,\]
 and Helmholtz free energy\[
F=U-\beta^{-1}S\]

We can define also the densities of the extensive (that is asymptotically
linear in $\Lambda$) thermodynamic variables in the thermodynamic
limit. For example we define the limiting Gibbs free energy for unit
volume as\[
g=\lim_{\Lambda\rightarrow\infty}\frac{G}{\Lambda}=\sum\mu_{j}c_{j}\]

\section{\char`\"{}Thermo\char`\"{} dynamics}

For given rate parameters $u,b,f,h$ the internal degrees of freedom
of the particles are independent and identically distributed random
variables. In other words, the distribution belongs to $\mathfrak{M}$.
However, the kinetic energies may have not $\chi^{2}$ distributions,
that is the velocities may not have Maxwell distribution. We will
force the kinetic energies to become $\chi^{2}$ using the limit $s_{f}\rightarrow\infty$.

Denote Gibbs state of the $j$-type ideal gas as $\mathcal{G}_{j}=\mathcal{G}_{j}(\beta,\mu_{j})$.
Define $\mathfrak{M}_{0}\subset\mathfrak{M}$ the set of all measures
$\times_{j}\mathcal{G}_{j}(\beta,\mu_{j})$ for any $\beta,\mu_{1},...,\mu_{J}$,
and $\mathfrak{M}_{0,\beta}$ - its subset with fixed $\beta$. In
physical and \ biological books on non-equilibrium thermodynamics,
there are some general statements, see for example \cite{TuKu}, which
hold for many concrete examples, in particular they will hold in our
model. Firstly, there is a submanifold in the space of probability
measures on the state space, defined by a finite number of macroparameters,
and moreover, this submanifold is invariant with respect to the full
dynamics. Secondly, each point of this manifold is a product of $k$
independent measures. In our case each point $\nu\in\mathfrak{M}_{0}$
is a product $\nu=\nu_{1}\times...\times\nu_{J}$ and the points of
$\mathfrak{M}_{0}$ are in one-to-one with the vector $\mathcal{M}=(\beta,\mu_{1},...,\mu_{J})$
of parameters. Note that a point of $\mathfrak{M}_{0}$ is also uniquely
defined by the vector $(\beta,c_{1},...,c_{J})$. The third general
statement concerns different time scales, that we discussed above.

The following result, in despite of its evidence, is crucial for discussing
chemical thermodynamics.

\begin{theorem}
The limits in distribution 
\begin{equation*}
\mathfrak{C}_{c}(t)=\lim_{s_{f}\rightarrow \infty }\mathfrak{X}_{c}(t),\mathfrak{O}_{c,\beta }(t)=\lim_{s_{h}\rightarrow \infty }\mathfrak{C}_{c}(t)
\end{equation*}exist for any fixed \(t\). Moreover, the manifold \(\mathfrak{M}_{0}\) is
invariant with respect to the process \(\mathfrak{C}_{c}(t)\) for any fixed
rates \(u,b,h\). The manifolds \(\mathfrak{M}_{0,\beta }\) are invariant with
respect to \(\mathfrak{O}_{c,\beta }(t)\).
\end{theorem}

Thus, in the process $\mathfrak{C}_{c}(t)$ the velocities have Maxwell
distribution at any time moment. For the process $\mathfrak{O}_{c,\beta}(t)$
moreover, at any time $t$ the temperature is equal to $\beta$, that
is there is heat exchange with the environment.

The resulting process $\mathfrak{C}_{c}(t)$ on $\mathfrak{M}_{0}$
can be defined, using the evolution of $c_{j}(t)$ and formula (\ref{chempot}),
also by the deterministic evolution of the vector $\mathcal{M}(t)=(\beta(t),\mu_{1}(t),...,\mu_{J}(t))$

\section{Thermodynamics of unimolecular reactions}

Let us make first some remarks about conserved quantities. Note that
$N=\sum_{j}<n_{j}>$ (for finite $\Lambda$) and $\sum_{j}c_{j}$
(for inifinite volume) are conserved. Then from the equation of state
(\ref{eqstate}) it follows that the pressure $P$ is conserved (for
fixed $\beta$), the same for the grand potential. Thus in our model
$N,P,\Lambda$ are conserved.

Further on we consider only the process $\mathfrak{O}_{c,\beta}(t)$.
Then all thermodynamic potentials are functions (for fixed $K_{1},...,K_{J}$)
on $\mathfrak{M}_{0\beta}$, for example the enthalpy $H$, or the
Gibbs free energy $G$.

\paragraph{Hess's law}

Consider two different processes $\mu_{j}^{(1)}(t)$ and $\mu_{j}^{(2)}(t),j=1,...,J$,
on $\mathfrak{M}_{0,\beta}$, for example with different reaction
rates. Assume also that for some $T>0$\[
\mu_{j}^{(1)}(0)=\mu_{j}^{(2)}(0),\mu_{j}^{(1)}(T)=\mu_{j}^{(2)}(T),j=1,...,J\]
 that is these two processes have the same initial and final points.
Then the Hess law says that the differences between initial and final
enthalpies are the same for both processes. In \ fact, this law holds
automatically in our model, because both processes are described by
two paths on $\mathfrak{M}_{0,\beta}$ with the same initial and final
points, and the enthalpy is a function on the invariant manifold $\mathfrak{M}_{0,\beta}$.

The simplest classification of reactions is in terms of the enthalpy
$H$. If $\Delta H=H(\infty)-H(0)<0$ then the reaction is called
exothermic, the heat $Q$ is goes to the environment, if $\Delta H>0$
the reaction is endothermic and the heat is taken from the environment.
That is $\Delta H=Q$.

\paragraph{Equilibrium conditions}

We assume further on that there are no slow binary reactions, moreover
we consider the case $J=2$. That is, consider the system with two
types and two reversible reactions $1\rightleftarrows2$. Thus we
have 2 parameters $\mu_{1},\mu_{2}$ and fixed $\beta$.

Let us remind how the equilibrium condition $\mu_{1}=\mu_{2}$ appears
in chemical thermodynamics. For the extensive variable $X=<n_{1}>$
the corresponding conjugate variable $A$ (thermodynamic force) is
(assuming $N=<n_{1}>+<n_{2}>$ fixed) called (chemical) affinity\[
A=-\frac{\partial G}{\partial X}|_{\beta,P,N}=-\mu_{1}+\mu_{2}=-\Delta G_{0}-\beta^{-1}\ln\frac{c_{1}}{c_{2}},\Delta G_{0}=\mu_{1,0}-\mu_{2,0}-(K_{1}+K_{2})\]
 $\Delta G_{0}$ is called the free energy of the reaction. Note that
instead of vectors $(\mu_{1},...,\mu_{J})$ for the points of $\mathfrak{M}_{0,\beta}$
one can use points $(c_{1},...,c_{J})$. Then $A$ can also be defined
as\[
A=-\frac{\partial g}{\partial c_{1}}|\beta,P,c\]
 The equation of state (relation between $X$ and $A$) is\[
c_{1}=\frac{c}{1+\exp(-\beta A-\Delta G_{0})}\]
 Equilibrium points are defined as points where $A=0$, this gives
$\mu_{1}=\mu_{2}$. From (\ref{chempot}) it follows that the equilibrium
condition $\mu_{1}=\mu_{2}$ in chemical thermodynamics uniquely defines
the quotient $\frac{c_{1,e}}{c_{2,e}}$ of the equilibrium densities
$c_{j,e}$. The equilibrium constant is defined as \begin{equation}
\kappa=\frac{c_{1,e}}{c_{2,e}}=\exp(-\beta\Delta G_{0})\label{equi_1}\end{equation}
 Moreover, for a given $c$ the equilibrium condition uniquely defines
a (fixed) point on $\mathfrak{M}_{0,\beta}$, that is the invariant
Gibbs measure.

\paragraph{Stochastic chemical kinetics reconstructed}

Now we give an example of such process in our case. Assume $K_{1}<K_{2}$.
Assume now the simplest possible dependence of $u_{jj^{\prime}}$
on $T$: $u_{jj^{\prime}}(T)$ equal some constants $w_{jj^{\prime}}$
if $T_{j}+K_{j}-K_{j^{\prime}}\geq0$, and $u_{jj^{\prime}}(T)=0$
otherwise. Then the process $\mathfrak{O}_{c,\beta}(t)$ can be given
explicitely. Denote $g_{\beta}(r)=P(\left|\xi\right|>r)$ for the
$\chi^{2}$ random variable $\xi$ with inverse temperature $\beta$.

It is easy to see that the process $\mathfrak{O}_{c,\beta}(t)$ can
be reduced to the Markov chain on $\left\{ 1,2\right\} $ with rates
\[
v_{21}=w_{21},v_{12}=g_{\beta}(K_{2}-K_{1})w_{12}\]

\paragraph{Monotonicity of Gibbs energy for fixed $\beta$}

This law says that Gibbs free energy $G$ has its minimum at the fixed
point and $G(t)$ is monotonic in time. It is evident in the vicinity
of the equilibrium point. One can say more, if the process $c_{j}(t)$
corresponds to some Markov process.

Let any Markov process with two states $1,2$ be given such that for
some constant $C$\begin{equation}
p_{1}(t)=Cc_{1}(t),p_{2}(t)=Cc_{2}(t),\pi_{1}=Cc_{1,e},\pi_{2}=Cc_{2,e}\label{process}\end{equation}
 where $p_{j}(t)$ are its probabilities at time $t$, and $\pi_{j}$
are its stationary probabilities.

Remind that for a finite irreducible Markov chain with the rates $w_{jj^{\prime}}$
the entropy\ of the positive measure $p=(p_{1},...,p_{J})$ relative
to the stationary measure $\pi=(\pi_{1},...,\pi_{J})$ is defined
as \begin{equation}
S_{M}=\sum p_{j}\ln\frac{p_{j}}{\pi_{j}}=C\sum c_{j}\ln\frac{c_{j}}{c_{j,e}}\label{Mark}\end{equation}

Now we will prove that the Gibbs free energy $g$ and Markov entropy
$S_{M}$ are equal up to a multiplicative and additive constants.

\begin{theorem}
At any time \(t\) we have for the Gibbs free energy density \(g(t)\)\begin{equation*}
g(t)=\mu c+\frac{1}{\beta C}S_{M}(t)
\end{equation*}where \(\mu =\mu _{1}=\mu _{2}\). It follows that \(g(t)\) is time monotone.

Moreover, the process \(p_{j}(t)\), satisfying (\ref{process}), is unique, up
to a common time scale.
\end{theorem}

Proof. For the Gibbs free energy density we get using (\ref{chempot})\begin{equation}
g=\lim_{\Lambda}\frac{G}{\Lambda}=\sum_{j}c_{j}\mu_{j}=\beta^{-1}\sum_{j}c_{j}\ln c_{j}+\sum_{j}c_{j}(\mu_{j,0}+K_{j})=\label{free_1}\end{equation}
 \[
=\beta^{-1}\sum_{j}c_{j}\ln c_{j}+\sum_{j}c_{j}(\mu-\beta^{-1}\ln c_{j,e})=\mu c+\beta^{-1}\sum_{j}c_{j}\ln\frac{c_{j}}{c_{j,e}}\]
 At the same time\[
S_{M}=\sum p_{j}\ln\frac{p_{j}}{\pi_{j}}=C\sum c_{j}\ln\frac{c_{j}}{c_{j,e}}\]
 As $S_{M}$ is known to decrease during Markov evolution, see \cite{Ligg},
the second assertion of the theorem follows as well.

Let us show now that there is unique choice of dynamics, that is of
the rates $v_{jj^{\prime}}$, which give equilibrium condition $\mu_{1}=\mu_{2}$.
Each Markov chain with two state is reversible, because reversibility
condition $\pi_{1}v_{12}=\pi_{2}v_{21}$ follows immediately from
Kolmogorov equation \[
\frac{d\pi_{1}}{dt}=\pi_{2}v_{21}-\pi_{1}v_{12}\]
 Then\begin{equation}
\frac{\pi_{1}}{\pi_{2}}=\frac{c_{1,e}}{c_{2,e}}\label{pic}\end{equation}
 In fact from\[
\frac{\pi_{1}}{\pi_{2}}=\frac{v_{21}}{v_{12}}\]
 and (\ref{pic}) it follows that $v_{jj^{\prime}}$ are uniquely
defined up to some constant $C$, which determines some common time
scale (speed of both reactions) and is irrelevant to thermodynamics.
Theorem is proved.

Relation with Onsager theory in our example is the following. The
flux is defined as\[
J_{1}=\dot{X}_{1}\]
 or in the thermodynamic limit\[
J_{1}=\frac{dc_{1}}{dt}\]
 And from the equations\[
\frac{dc_{1}}{dt}=c_{2}u_{21}-c_{1}u_{12},c_{2}=c-c_{1}\]
 we have\[
J_{1}=\frac{1-\exp(-\beta A)}{u_{21}^{-1}+u_{12}^{-1}\exp(-\beta A)}\]

\paragraph{Energy redistribution}

Assume that at time $t=0$ an arbitrary distribution $p_{0}(j,T)$
of the vector $(j,T)$ is given. Then at any $t>0$ the densities
$p_{t}(j,T)$ for any particle will be \begin{equation}
c\sqrt{T}\exp(-\beta T)p_{t}(j)\label{prod1}\end{equation}
 for some $p_{t}(j)$. This can be shown as follows. As the internal
degrees of freedom of infinite number of particles are i.i.d. random
variables, then there exist a.s. the limits\[
\bar{T}(t)=\lim_{\Lambda\rightarrow\infty}\frac{1}{\Lambda}\sum_{i:x_{i}\in\Lambda}T_{i}(t),\bar{K}(t)=\lim_{\Lambda\rightarrow\infty}\frac{1}{\Lambda}\sum_{i:x_{i}\in\Lambda}K_{j_{i}}(t)\]
 exist at any time $t$. In particular, a.s. for any fixed values
of $K_{j_{i}}(t)$ the limits\[
\bar{T}(t)=\lim_{\Lambda\rightarrow\infty}\frac{1}{\Lambda}\sum_{i:x_{i}\in\Lambda}T_{i}(t,\vec{K}(t))\]
 exist and are equal. Here $\vec{K}(t)=\left\{ K_{j_{i}}(t),i=1,2,...\right\} $.

Moreover for any given $\vec{K}(t)$ there is a sequence of jump moments\[
t_{1}<t_{2}<...<t_{n}<...\]
 of fast binary collisions and heat transfer, which do not change
parameters $j_{i}$ (and thus $K_{j_{i}}$) of the molecules. If $s_{f}$
and $s_{\beta}$.tend to infinity we have a.s. there will be \char`\"{}infinite\char`\"{}
number of fast collisions and heat transfers between any two unary
reactions. It follows that any time $t$ we have a product measure
(\ref{prod1}).

We will study the sequence $\bar{K}(t)$. As any time moment $t\geq0$
we have $\bar{T}(t)=\beta^{-1}$ put also $\bar{T}(0)=\beta^{-1}$
for continuity. Now there two possibilities:

\begin{enumerate}
\item $\bar{K}(0)<\bar{K}(\infty)$, this means that the kinetic energy,
pumped up to the system with the heat, is transformed to the chemical
energy; 
\item $\bar{K}(0)>\bar{K}(\infty)$, this means that the chemical energy
is transformed to the kinetic energy, which goes out as the heat. 
\end{enumerate}

\section{Further Problems}

This paper is a kind of advertisement for mixed dynamics. Pure local
dynamics, even in one dimension, leads immediately to too difficult
problems. Mixed dynamics is simpler and many situations could be modelled
with it, especially in biology. It is quite natural to discuss here
possible related problems, there are many.

\paragraph{Logical structure}

From one side, chemical thermodynamics has some distinct logical structure,
from the other side this structure is based on some approximations.
Our model suggests a distinct implementation of this logical picture,
and shows what is the nature of the approximations. Moreover, there
are fundamental questions. We go now to more detailed discussion:

\begin{itemize}
\item Chemical thermodynamics largely uses ideal gas formulas, for example
see formula (\ref{chempot}). For this reason the corresponding expressions
can be only approximate; 
\item Equilibrium conditions play the central role in the chemical kinetics.
In fact, they are based on the assumption that the chemical equilibrium
corresponds to the minimum of the Gibbs free energy, in a sufficiently
large class of measures, see the end of this section. It is not at
all clear for me whether this should be considered as a fundamental
experimental fact or it should be deduced from microscopic dynamics.
A possible key could be the coincidence of some thermodynamic potential
with Lyapounov function for the dynamics, see the above example. See
also \cite{Maes,LebMae}; 
\item The dynamics for a system with chemical reactions is ambiguous itself.
The reactions can be incorporated into hamiltonian dynamics only via
some probabilistic mechanism. It is what we do here, using another
field of physical chemistry - stochastic chemical kinetics. This dynamics
cannot be arbitrary - the constraints on it are posed by the equilibrium
conditions, given apriori. 
\item There is also a deeper reason for the dynamics ambiguity. If we do
not want to use random mechanisms for reaction, we are encountered
with the dual nature of bound states. From one side, bound states
are considered (in chemical thermodynamics) as fundamental particles
at EACH (except discrete time moments when reactions occur) time moment.
From the other side, it appears as a composite particle (in the classical
physics) from hamiltonian dynamics via scattering theory. In the latter
for the\ bound state formation one needs INFINITE or at least finite
time interval. Thus, it is ambiguous to prescribe when the new composite
particle appears. 
\item The same problems arise for quantum hamiltonians with chemical reactions,
in terms of annihilation-creation operators, with non-quadratic terms
corresponding to collisions. 
\item Possibly there is some escape from all these problems even in the
general local models, that is for nonideal gases with interaction
between different gases. There should be equivalent representation
of this complex system by ideal gases of quasiparticles. The corresponding
quasiparticles could even correspond to real particles surrounded
with clouds, that is the particles become slightly renormalized. However,
quasiparticle representation can be obtained now rigorously only for
some ground state models, and only for equilibrium dynamics, see \cite{MalMin}.
This approach brings us to another \textit{tabula rasa}: consider
an infinite particle system where elementary particles are atoms,
not molecules. Then we are in the framework of purely Hamiltonian
system. One should be able to show that the dynamics brings this measure
to the configurations where most atoms form bound states - molecules. 
\end{itemize}

\paragraph{Non ideal systems}

The deterministic part of the theory of chemical networks is presented
in \cite{Oth}, in completely rigorous beautiful framework. However,
there was no energy component, no probability and no microscopic dynamics.

The logical framework of \cite{Oth} is the following: deterministic
chemical kinetics is postulated together with some restrictions on
the invariant manifolds, related to the (also postulated) Gibbs free
energy $G$. It is presented as\[
\frac{\partial G}{\partial c_{j}}=\mu_{j}=\mu_{j,0}(\beta,P)+\beta^{-1}\ln\gamma_{j}(c)c_{j}\]
 where $\gamma_{j}(c_{j})$ are some unknown functions of $c_{j}$.
If $\gamma_{j}(c_{j})=c_{{}}^{-1}$ for all $j$, then the system
is called ideal. As for the nonideal systems, microscopic models should
give information about $\gamma_{j}$.

However, even for nonideal system the same question as above will
be the main enigma of the chemical thermodynamics.

\paragraph{More thermodynamical processes}

We did not consider chemical thermodynamics for binary reactions in
this paper. However, it is clear that it can be done, because (as
it is shown in section 2) its inifinite particle dynamics is quite
similar to unimolecular dynamics. Also reactions which do not conserve
$N$ are of interest. In particular, decay and synthesis that is $A\rightarrow B+C$
and $A+B\rightarrow C$. Here for the first reaction one should assign
somehow the coordinate to $B$ and $C$. It can be done in the following
way: one molecule, for example $B$, with probability $\frac{1}{2}$
will have the coordinate of $A$, then $C$ is put randomly into $\Lambda$.
It seems unnatural in a finite volume, but in the infinite volume,
it will give, as for slow binary reaction, a local process for particles.
Together with evolution of densities.

We are lacking microscopic models even for simpler situations in non-equilibrium
thermodynamics: local models quickly become too difficult to be useful.
However mixture of local models with mean field dynamics looks quite
promising, and tractable. For example, one could consider exchange
of matter with the environment, work and efficiency produced by mechanochemical
and chemochemical machines, etc., see \cite{MaeWie}.

In quantum case there can be other statistics, Fermi and Bose, reactions
with them are interesting to consider. Also one could try to model
reactions in solutions or reactions with large $P$, nuclear reactions
etc. Some substitutes for Clausius entropy are used in nuclear physics,
for which there are no dynamical models.

\paragraph{Biology}

In biology it is known heuristically that the chemical networks may
have different time scales. First scale is the fundamental microscale,
it is the fastest scale, where local equilibrium establishes for some
thermodynamic parameters (for our model it was the global equilibrium).
Second scale (call it micro non-equilibrium), is the scale of main
concrete reactions.

If the chemical network is large enough there can be also other scales,
even slower than the second one. For example, genetic networks can
be modelled as if the list of reactions changes with time, slower
than the scale of the reactions.

It seems very important to understand and classify these scales and
model all main time scales. One cannot yet even pose exact mathematically
reasonable questions here.

\paragraph{Variational Problems }

Assume that some system has states $1,2,...$ with energy levels $\varepsilon_{k}$
of the state $k$. Gibbs distribution on the set $\left\{ 1,2,...\right\} $
is defined as\[
p_{k}=Z^{-1}\exp(-\beta\varepsilon_{k}),Z=\sum_{k}\exp(-\beta\varepsilon_{k})\]
 Then it is known and easy to show that Gibbs equilibrium state is
the state of maximum entropy $S$ for fixed mean energy $U$. To see
this we are looking for extrema of\[
S=-\sum_{k}p_{k}\ln p_{k}+\lambda\sum_{k}\varepsilon_{k}p_{k}\]
 with two constraints\[
U=\sum_{k}\varepsilon_{k}p_{k}=c,\sum_{k}p_{k}=1\]
 Thus we are looking for extrema of\[
-\sum_{k}p_{k}\ln p_{k}+\lambda\sum_{k}\varepsilon_{k}p_{k}+\gamma\sum_{k}p_{k}\]
 Differentiation gives\[
-\ln p_{k}+1+\lambda\varepsilon_{k}+\gamma=0\]
 That is\[
p_{k}=C\exp\lambda\varepsilon_{k}\]
 where $\lambda<0$ for convergence reason.

Similarly, equilibrium state is the state of minimum mean energy $U$
for fixed entropy $S$. Here we differentiate\[
U+\lambda S+\gamma\sum_{k}p_{k}=\sum_{k}\varepsilon_{k}p_{k}-\lambda\sum_{k}p_{k}\ln p_{k}+\gamma\sum_{k}p_{k}\]
 Grand canonical ensemble is included to the previous scheme. In fact,
\ consider grand canonical ensemble\[
\sum_{N=0}^{\infty}\exp\beta((\mu N-\sum_{k=1}^{\infty}\varepsilon_{Nk})\]
 where $\varepsilon_{Nk}$ are the energies levels of the system with
$N$ particles. This case can be reduced to the previous one if $\mu N$
is considered among the energy levels, that is introduce $\varepsilon_{N0}=-\mu N$.

The Helmholtz free energy $A=U-\beta^{-1}S$ is defined for any measure,
that is for any system $\left\{ p_{k},\varepsilon_{k}\right\} $.
In our case $P$ and $\Lambda$ are constant, as \[
P=\beta^{-1}c\]
 Thus the Gibbs free energy $G=A+P\Lambda$ is also defined for some
class of measures, including our manifold $\mathfrak{M}_{0}$. It
could be interesting to know the widest class of measures, where $G$
is defined and is a Lyapounov function for an appropriate \char`\"{}thermo\char`\"{}
dynamics.

\section*{Appendix}

Here we present some technicalities omitted in the main part of the
paper. We use here natural and intuitive, but {}``not quite standard'',
approach to the convergence proof of mean field type Markov processes
with large number of particles. It does not use standard techniques
of martingale problem, semigroup generators, tightness of measures,
etc. This approach, coming from cluster expansion ideas, is based
on small piece of combinatorics and simple probabilistic estimates.
To avoid cumbersome notation, we present this method for binary reactions
only, that is general enough to see all peculiarities. Moreover, it
is clear that this approach can be applied to many other situations
as well. This techniques proves both smooth dependence of the limiting
distributions on $t$ and on initial data, and the chaos property.

\paragraph*{The model}

Consider continuous time Markov process $\xi^{N}(t)=(\xi_{v}^{N}(t),v=1,...,N),t\in[0,\infty)$
with state space $S^{N}$, where $S$ is some space of one-particle
states. To define the process we fix some linear operator $U:M(S^{2})\rightarrow M(S^{2})$,
where $M(S^{2})$ is the set of measures on $S^{2}$ with variation
norm. This operator is defined by the conditional measures $d\sigma(s_{1},s_{1}^{\prime}|s,s^{\prime})$,
where $d\sigma$ is the family, indexed by the pairs $(s,s^{\prime})$,
of probability measures on the set of pairs $(s_{1},s_{1}^{\prime})$.
For any pairs $v,v^{\prime}$ of different particles the operators
$U(v,v^{\prime})$ define transformations on the set $M(S^{N})$ of
measures on the state space $S^{N}$. This operator acts as $U$ only
on $v$ and $v^{\prime}$ components of $S^{N}$, it is assumed symmetric
with respect to permutation $v\leftrightarrow v'$.

The transitions of the process are defined in two steps. Firstly,
each particle $v=1,...,N$ generates independent Poisson process of
time moments with rate $\lambda$. Denote $N_{v}=N_{v}(\omega)$ the
random number of time moments on the time interval $[0,t]$, generated
by the particle $v$ and let\[
0<t_{v,1}(\omega)<...<t_{v,N_{v}(\omega)}(\omega)<t\]
 be the these moments. We have\[
P(N_{v}(\omega)=n)=\frac{(\lambda t)^{n}}{n!}exp(-\lambda t)\]
 and the density of the vector $(t_{1}<...<t_{n})$ is given by\[
p(t_{1}<...<t_{n})dt_{1}...dt_{n}=
\]
\[ = exp(-\lambda t_{1})\lambda dt_{1}...exp(-\lambda(t_{n}-t_{n-1}))\lambda dt_{n}exp(-\lambda(t-t_{n}))=\]
 \[
=exp(-\lambda t)\lambda^{n}dt_{1}...dt_{n}\]
 In other words, it is uniform on the simplex $\{0<t_{1}<...<t_{n}<t\}$.
At each time moment $t_{v.i}$ a pair $(v,w_{i})$ is produced, where
$w_{i}\neq v$ is chosen with probabiity $\frac{1}{N-1}$. One could
say equivalently that each pair generates independent Poisson process
with rate $\frac{2\lambda}{N-1}$. Merging them together, we get the
combined Poisson process, which we denote\[
0<t_{1}(\omega)<t_{2}(\omega)<...<t_{n}(\omega)<t\]
 It has rate (density) $N\lambda$.

For a given $\omega$ denote $v(i,\omega),v^{\prime}(i,\omega)$ the
pair produced at time $t_{i}$ in the combined process. In fact, we
can take $\omega$ as the sequence\begin{equation}
(t_{i},v_{i},v_{i}^{\prime}),i=1,2,...\label{omega}\end{equation}
 itself.

We will consider (ordered) sequences of (unordered) pairs\begin{equation}
\theta=(v_{1},w_{1}),...,(v_{n},w_{n})\label{theta}\end{equation}
 We call $\left|\theta\right|=n$ the length of $\theta$. Thus for
any $\omega$ the sequence\[
\theta=\theta(\omega)=((v_{i},v_{i}^{\prime}),i=1,2,...)\]
 is defined.

For any sequence $\theta$ we define the chronological product of
measure transformations\[
U(\theta)=U(v_{n},w_{n})...U(v_{1},w_{1})\]
 acting from left to right. Thus, for any $\omega$ the quantities
$\theta(\omega),n=n(\omega)$ and $U(\omega)=U(\theta(\omega))$ are
uniquely defined. The measure on $S^{N}$ at time $t$ is given by\[
\mu^{N}(t)=\int U(\theta(\omega))d^{N}\omega\]
 where $d^{N}\omega$ is the measure on Poisson trajectories for given
$N$.

Assume, for any $N$, that at time $t=0$ the one-particle distributions
$\mu_{v}(0)$ of $s_{v}$ are i.i.d. Then we have the following result.

\begin{description}
\item [Theorem]For any $v=1,2,...$ one-particle measures $\mu_{v}^{N}(t)$,
converge as $N\rightarrow\infty$ uniformly in $\mu_{v}(0)$ on any
finite interval $(0,\tau_{0})$, to some $\mu_{v}(t)$, identical
for any $v$. Moreover, $k$-particle distributions $\mu_{1,...k}^{N}(t)$
converge to the product $\mu_{1}(t)\times...\times\mu_{1}(t)$. 
\end{description}
The plan of the proof will be the following. For $\tau_{0}$ sufficiently
small we will obtain explicit absolutely convergent series for any
finite-dimensional distribution. This gives complete control for small
times. To prove the same properties for larger $t$ one can use the
semigroup property (as one can write $t=k\tau_{0}+t^{\prime}$ for
some $k$ and $t^{\prime}<\tau_{0}$) and uniformness on the initial
one-particle distribution.

\paragraph*{Clusters}

We need some combinatorics. We call abstract sequence (\ref{theta})
of pairs connected if $V_{k}\bigcap\{ v_{k},w_{k}\}\neq\varnothing$
for any $k=1,...,n-1$, where \[
V_{k}=\bigcup_{i=k+1}^{n}\{ v_{i},w_{i}\}\]
 The pair $\{ v_{i},w_{i}\}$ in $\theta$ is called essential if
$\{ v_{i},w_{i}\}$ does not belong to $V_{i}$. Connected sequence
is called essential if all its pairs are essential. It follows that
there are no identical pairs in the essential connected sequence.
We call $w$-sequence any connected sequence with $w\in\{ v_{n},w_{n}\}$.

For any $v$ and any $\omega$ we define a subsequence $\theta_{v}(\omega)$
of $\theta(\omega)$ as follows. Take maximal $j$ such that $v\in\left\{ v_{j},w_{j}\right\} $.
If there is no such $j$ then put $\theta_{v}(\omega)=\varnothing$.
If there is such $j=j(\omega)$ then we define $\theta_{v}(\omega)$
as the minimal subsequence of $\theta(\omega)$ satisfying the following
two conditions: 1) if $v\in\{ v_{i},w_{i}\}$ then $\{ v_{i},w_{i}\}\in\theta_{v}(\omega)$;
2) if for $i\leq j$ the pair $\left\{ v_{i},w_{i}\right\} $ belongs
to $\theta_{v}(\omega)$, then any pair $\left\{ v_{k},w_{k}\right\} $
for $k<i$ and such that $\left\{ v_{i},w_{i}\right\} \cap\left\{ v_{k},w_{k}\right\} \neq\varnothing$,
also belongs to $\theta_{v}(\omega)$.

\paragraph{Resummation formula}

For given $v$ and any connected $v$-sequence $\theta$ we consider
the probabilities\[
P^{N}(\theta)=P(\theta_{v}(\omega)=\theta),P_{k}^{N}=\sum_{\left|\theta\right|=k}P^{N}(\theta)\]
 and will use the following resummation formula\[
\mu_{1}^{N}(t)=B_{1}\sum_{\theta}P^{N}(\theta)U(\theta)(\times_{v=1}^{l_{n}(\theta)}\mu_{v}(0))\]
 where $B_{1}$ is the projection on the distribution of particle
$1$, $\times_{v=1}^{l_{n}(\theta)}\mu_{v}(0)$ is the initial distribution
on $S^{\left|\theta\right|}$ and $l_{k}(\theta)$ is the number of
elements in the union $\bigcup_{i=n-k+1}^{n}\{ v_{i},w_{i}\}$.

It is easy to get explicit formula for $P_{k}^{N}$. For example,
for $n=0$ that is for the empty $\theta$ we get $P_{0}^{N}=\exp(-2\lambda t)$,
for $n=1$ we have \[
P_{1}^{N}=2\lambda\int\exp(-2\lambda(t-t_{1}))\exp(-2\lambda(2-\frac{1}{N-1})t_{1})dt_{1}\]
 that is there is no particle $1$ on the time interval $(t_{1},t)$
and there are no particles $1$ and $i$ (assuming that $\theta=(1,i)$)
on the interval $(0,t_{1})$. For $n=2$ we have \[
P_{2}^{N}=\sum_{i\neq j}P(\theta_{1}(\omega)=(1,i),(1,j))+\sum_{i}P(\theta_{1}(\omega)=(1,i),(1,i))+\]
 \[
+\sum_{i\neq j}P(\theta_{1}(\omega)=(1,i),(i,j))\]
 where for example,\[
\sum_{i\neq j}P(\theta_{1}(\omega)=(1,i),(1,j))=\]
 \[
=(2\lambda)^{2}\frac{N-2}{N-1}\int\int\exp(-2\lambda(t-t_{2}))\exp(-2\lambda(2-\frac{1}{N-1})(t_{2}-t_{1})) \times
\]
\[
\times \exp(-\frac{2\lambda}{N-1}a_{3}t_{1})dt_{1}dt_{2}\]
 where $a_{3}$ is the number of pairs, which intersect with $\{1,2,3\}$.
In the general case the formula looks quite similar\[
P^{N}(\theta)=(\frac{2\lambda}{N-1})^{n}\int...\int\prod_{k=1}^{n}\exp(-m(N,k,\theta)\frac{2\lambda}{N-1}(t_{k}-t_{k-1}))dt_{1}...dt_{n}\]
 where $m(N,k,\theta)$ are some positive numbers. However, we will
simplify our task: we do not need exact expression for the exponents
because we will use the estimates\[
\exp(-m\lambda(t_{k}-t_{k-1}))\leq1,\left\Vert U(\chi)\right\Vert \leq1\]
 For the integration we will use\begin{equation}
\int...\int dt_{1}...dt_{n}=\frac{t^{n}}{n!}\label{simplex}\end{equation}
 It is convenient to consider equivalence classes of connected sequences.
We say that two sequences are equivalent if one can be obtained from
the other by some permutation $\phi:\{1,...,N\}\rightarrow\{1,...,N\}$
of particles. Note that $m(N,k,\theta)$ and $P^{N}(\theta)$ depend
only on the equivalence class.

Let $A$ be a subset of $\{1,...,n\}$. We say that a sequence $\theta$
has type $A$ if the pair $\{ v_{i},w_{i}\}$ is essential for any
$i\in\{1,...,n\}\setminus A$ and nonessential otherwise. Essential
sequences correspond to $A=\varnothing$. The number $C_{n,ess}^{(N)}$
of essential sequences of length $n$ satisfies the following properties
\begin{equation}
(\frac{1}{N-1})^{n}C_{n,ess}^{(N)}=\prod_{k=1}^{n}k\frac{N-k}{N-1}\leq n!,(\frac{1}{N-1})^{n}C_{n,ess}^{(N)}\rightarrow n!\label{ess}\end{equation}
 For any nonempty $A$ we get similarly\begin{equation}
(\frac{1}{N-1})^{n}C_{n,A}^{(N)}\leq n!,(\frac{1}{N-1})^{n}C_{n,A}^{(N)}\rightarrow0\label{noness}\end{equation}
 as $N\rightarrow\infty$.

We need also the evident property that $m(N,k,\theta)\frac{1}{N-1}$
tend to some $m(k,\theta)$ as $N\rightarrow\infty$. Finally we have\[
\mu_{1}^{N}(t)=B_{1}\sum_{n}(\frac{2\lambda}{N-1})^{n}\sum_{\theta:|\theta|=n}\int...\int\exp(-2\lambda m(k,\theta)(t_{k}-t_{k-1})) 
\]
\[
\times dt_{1}...dt_{n}\sum_{ess}U(\theta)(\mu_{1}(0)\times...\times\mu_{l_{n}(\theta)}(0))\]
 By (\ref{noness}), in the limit non essential sequences do not count
and we get using (\ref{ess}) \[
\mu_{1}(t)=B_{1}\sum_{n}\lambda^{n}\sum_{\theta:|\theta|=n}\int...\int\exp(-2\lambda m(k,\theta)(t_{k}-t_{k-1}))
\]
\[
\times dt_{1}...dt_{n}\sum_{ess}U(\chi)(\mu_{1}(0)\times...\times\mu_{n+1}(0))\]
 For small $t$ the terms of both series have uniform exponentiall
bounds by (\ref{simplex}). Moreover, due a term-by-term convergence,
\[
\mu_{1}(t)-\mu_{1}^{N}(t)\]
 tends to zero in the norm. $C^{\infty}$ dependence on $t$ follows
from this. Chaos property can be shown quite similarly.

\end{document}